\documentclass[%
 aip,
 sd,%
 amsmath,amssymb,
 reprint, twocolumn,
]{revtex4-1}

\usepackage{graphicx}% Include figure files
\usepackage{dcolumn}% Align table columns on decimal point
\usepackage{bm}% bold math
\begin{document}

\preprint{AIP/123-QED}

\title{Influence of a charge-gradient force on dust acoustic waves}% Force line breaks with \\
%\thanks{Footnote to title of article.}

\author{Alexey G. Khrapak}
 \email{khrapak@mail.ru}
 \affiliation{Joint Institute for High Temperatures, Russian Academy
of Sciences, 125412 Moscow, Russia}%Lines break automatically or can be forced with \\

\author{Sergey A. Khrapak}%
 %\email{Second.Author@institution.edu.}
\affiliation{Aix-Marseille-University, CNRS, PIIM, 13397 Marseille, France%\\This line break forced with \textbackslash\textbackslash
}%
\affiliation{Institut f\"{u}r Materialphysik im Weltraum, Deutsches Zentrum f\"{u}r Luft- und Raumfahrt (DLR), D-82234 We{\ss}ling, Germany}
%\altaffiliation[Also at ]{Physics Department, XYZ University.}

% \homepage{http://www.Second.institution.edu/~Charlie.Author.}
%\affiliation{%
%Second institution and/or address%\\This line break forced% with \\
%}%

\date{\today}% It is always \today, today,

\begin{abstract}
The influence of a charge-gradient force, associated with variations of the particle charge in response to external perturbations, on the propagation of low-frequency waves in weakly coupled complex (dusty) plasmas is investigated. The magnitude of the effect is compared with that due to polarization force, studied previously in the literature. Numerical estimates are presented for the regime, where the orbital motion limited approach to particle charging is relevant.
\end{abstract}

%\pacs{52.27.Lw, 52.35.Dm}% PACS, the Physics and Astronomy
\maketitle

A complex plasma represents an ionized gas containing electrons, ions, neutral atoms or molecules, and massive dust particles. The charged dust grains embedded into a plasma not only change the electron--ion composition and thus affect conventional wave modes (e.g., ion--acoustic waves), but also introduce new low-frequency modes associated with the microparticle motion, alter dissipation rates, give rise to instabilities, etc.~\cite{TsytovichUFN,FortovUFN,ShuklaRMP} Moreover, the particle charges vary in time and space, resulting in important qualitative differences between complex plasmas and usual multi-component plasmas.\cite{Fortov05,MorfillRMP} The focus of this brief communication  is on the influence of the plasma background and grain charge variability on linear waves in weakly coupled unmagnetized complex plasma.

In the long-wavelength limit, collective excitations of the particle component exhibit acoustic-like dispersion and are therefore called the ``dust acoustic waves'' (DAWs). The dispersion relation of DAWs for an ideal isotropic complex plasma was originally derived by Rao, Shukla, and Yu.\cite{Rao90}
In the original derivation of the DAW dispersion relation, a simplest fluid description of multicomponent plasmas was used. Several important effects were neglected, including  charge variations and specific forces acting on the charged particles (such as, for example, ion, electron, and neutral drag forces). One of the forces which can affect particle transport (also neglected originally) is the so called ``polarization'' force, discussed by Hamaguchi and Farouki.\cite{HamaguchiPRE1994,HamaguchiPoP1994}  This force was originally related to the presence of the density gradient in the plasma surrounding the particle.  In most practical cases the polarization force is small as compared to other forces present in the system. However, it was pointed out later that the polarization force can significantly affect propagation of the dust acoustic linear and non-linear waves.\cite{KhrapakPRL2009,BandiNJP2010,BandiPoP2012} This topic is presently under active investigation, for some relevant examples see Refs.~\onlinecite{MerlinoPPCF2012,PrajapatiPLA2015,SharmaEPL2016,BentadetIEEE2017} and references therein.

Recently, it has been demonstrated that the polarization force can contain a term proportional to the gradient of the particle charge, if the charge is not assumed fixed.\cite{KhrapakPRE03_2015} The derivation is straightforward.  The energy of an individual point-like test charge $Q$ immersed in an ideal plasma is
\begin{equation}\label{U}
 U=\frac{Q}{2}\left[\phi(r)-\frac{Q}{r}\right]_{r\rightarrow 0}=-\frac{Q^2}{2\lambda_{\text D}},
\end{equation}
where $\phi(r)=Q\exp(-r/\lambda_{\rm D})/r$ is the screened Coulomb (Debye-H\"uckel) potential,  $\lambda_D$ is the linearized Debye radius, $\lambda_{\text D}=\lambda_{{\text D}i}/\sqrt{1+(\lambda_{{\text D}i}/\lambda_{{\rm D}e})^2}$,
$\lambda_{{\rm D}i(e)}=\sqrt{T_{i(e)}/4\pi e^2n_{i(e)}}$,
and $T_{i(e)}$ and $n_{i(e)}$ are ion (electron) temperature (expressed in energy units) and density, respectively. If the charge is constant and the plasma is non-uniform, the particle will be acted by the force ($F=-\nabla U$)
\begin{equation}\label{F1}
F_{\rm pol}= -\frac{Q^2}{2}\frac{\nabla\lambda_{\text D}}{\lambda_{\text D}^2},
\end{equation}
which is known as the polarization force.\cite{HamaguchiPRE1994,HamaguchiPoP1994} It pushes the particles into the region where the Debye radius is smaller (that is where the temperature is lower and/or plasma density is higher).  If the charge is allowed to vary there is another contribution to the force
 \begin{equation}\label{F2}
  F_Q=\frac{Q\nabla Q}{\lambda_D}.
\end{equation}
This force is proportional to the gradient of the particle charge and we call it in the following the ``charge-gradient force''. The charge-gradient (CG) force pushes positively (negatively) charged particles to the region where their charge is higher (lower).  The purpose of this work is to report on the effect of this charge-gradient (CG) force on the linear dust acoustic waves. In particular, we will be interested in its relative magnitude, as compared to the conventional polarization force.

In the following we consider the most simple situation in order to single out the effects associated with the polarization and charge-gradient forces. We neglect all processes that can be neglected in this study and employ all reasonable simplifications. The consideration is to some extent similar to that of Ref.~\onlinecite{KhrapakPRL2009}.

The particle component is described by the continuity and momentum equations:
\begin{equation}\label{Nd}
  \frac{\partial n_d}{\partial t}+\nabla(n_d\textbf{v}_d)=0,
\end{equation}
\begin{equation}\label{Vd}
  \frac{\partial \textbf{v}_d}{\partial t}+(\textbf{v}_d\cdot\nabla)\textbf{v}_d=-\frac{Q}{m_d}\nabla\varphi+\frac{\textbf{F}_{\Sigma}}{m_d},
\end{equation}
where  $\textbf{v}_d$ and $m_d$ are the grain velocity and mass, $\varphi$ is the potential of the electric field acting on the particles, and ${\bf F}_{\Sigma}$ is the sum of all other forces. For the sake of simplicity below we consider only the polarization (\ref{F1}) and charge-gradient (\ref{F2}) forces. Note, that we have also omitted pressure term in Eq.~(\ref{Vd}). We further assume that the wave propagation results in small perturbations, $n_a=n_{a0}+n_{a1}$ $(a=e,i,d)$, $Q=Q_0+Q_1$, $\varphi=\varphi_1$,  $\textbf{v}_{d}=\textbf{v}_{d1}$, etc. If the perturbations are small (linear regime), the densities of electrons and ions satisfy the Boltzmann relations~\cite{Alexandrov84,Fortov00}
\begin{equation}\label{Nei}
  n_{i1}=-n_{i0}\frac{e\varphi_1}{T_i},\qquad n_{e1}=n_{e0}\frac{e\varphi_1}{T_e}.
\end{equation}
In the long-wavelength limit (where the dispersion relation is  acoustic) the densities of charged components satisfy the charge neutrality condition
\begin{equation}\label{P1}
 n_i-n_e+Zn_d =0,
\end{equation}
where $Z=Q/e$ is the particle charge number (note that $Z$ is negative for a negatively charged particle). Since the particle charge is not fixed, the system should be supplemented by the charging equation. In a rather  general form, the charging equation is~\cite{FortovUFN}
\begin{equation}\label{charge1}
\frac{\partial Z_1}{\partial t}+\Omega_{\rm ch}Z_1=J_0\left(\frac{n_{i1}}{n_{i0}} - \frac{n_{e1}}{n_{e0}} \right),
\end{equation}
where $Z_1$ is a variation of the particle charge number, $\Omega_{\rm ch}$ is the characteristic charging frequency and $J_0$ is the equilibrium flux of ions/electrons that the particle collects from the surrounding plasma. The equilibrium charge $Q$, associated with the equilibrium (floating) surface potential of the particle, is determined from the flux balance condition $J_i=J_e=J_0$. Quite generally, particle charging in a plasma is a very fast process~\cite{FortovUFN,Fortov05} and its characteristic frequency scale is much higher than frequency scales related to particle dynamics (e.g. DAW frequency scale). Therefore, we can write
\begin{equation}\label{charge2}
Z_1=\frac{J_0}{\Omega_{\rm ch}}\left(\frac{n_{i1}}{n_{i0}} - \frac{n_{e1}}{n_{e0}} \right).
\end{equation}

The system of equations (\ref{Nd}) -- (\ref{charge2}) is linearized following a standard procedure, i.e. assuming the $\sim \exp(i\textbf{kr}-i\omega t)$ dependence for all perturbations. In addition, we make one more simplification assuming that the electron temperature is much higher than the ion temperature, as it is in most complex plasmas occurring in gas discharges.
This implies  $\lambda_{{\rm D}e}\gg \lambda_{{\rm D}i}$, that is  $\lambda_{\rm D}\simeq \lambda_{{\rm D}i}$, and $|n_{i1}/n_{i0}|\gg |n_{e1}/n_{e0}|$.
After some simple algebra we obtain the dispersion relation of the form
\begin{equation}\label{DR}
\omega^2\left(1+\frac{n_{d0}}{n_{i0}}\frac{J_0}{\Omega_{\rm ch}}\right)=\omega_{d}^2\lambda_{\rm D}^2k^2\left(1+{\mathcal R}_{\rm pol}+{\mathcal R}_{\rm Q}\right),  	
\end{equation}
where
\begin{equation}\label{R1}
{\mathcal R}_{\rm pol}= \frac{Qe}{4\lambda_{\rm D}T_i}
\end{equation}
and
\begin{equation}\label{R2}
{\mathcal R}_{\rm Q}= \frac{J_0e^2}{\lambda_{\rm D}\Omega_{\rm ch}T_i}.
\end{equation}

Equation (\ref{DR}) represents the long-wavelength dispersion relation of low-frequency waves in the considered system. It is the main result of this study, its detailed analysis will follow.

First, if charge variations are neglected and the particle charge is fixed, which corresponds to the formal limit $\Omega_{\rm ch}\rightarrow \infty$, the dispersion relation is reduced to
\begin{equation}\label{DR1}
\omega^2=\omega_{d}^2\lambda_{\rm D}^2k^2\left(1+{\mathcal R}_{\rm pol}\right),  	
\end{equation}
which essentially coincides with the long-wavelength limit of the expression derived in Ref.~\onlinecite{KhrapakPRL2009}. Since the particle charge is usually negative in gas discharges, the quantity ${\mathcal R}_{\rm pol}$ is also negative. Thus, the actual dust-acoustic (sound) velocity is reduced compared to the conventional DAW velocity ($C_{\rm DAW}=\omega_{d}\lambda_{\rm D}$) by a factor $\sqrt{1+{\mathcal R}_{\rm pol}}$. For very large grains $|{\mathcal R}_{\rm pol}|$  can even approach  unity. In this case the net force on the grains is no longer a restoring force, and then the dispersion relation (\ref{DR1}) admits a transition from propagating DA waves to aperiodically growing perturbations. The effect of the charge-gradient force is expressed by the term ${\mathcal R}_{\rm Q}$ in the right-hand side of Eq.~(\ref{DR}). This term is obviously positive and thus it {\it reduces} the effects associated with the polarization force. The directions of the forces can also be verified as follows. For the polarization component we have
\begin{displaymath}
\textbf{F}_{\rm pol} \propto -\nabla\lambda_{\rm D}\propto\nabla n_i \propto -\nabla\varphi,
\end{displaymath}
and it acts in the direction of the electric field. For the CG component we have on the other hand
\begin{displaymath}
\textbf{F}_{\rm Q} \propto Q\nabla Q \propto Q\nabla n_i \propto -Q\nabla\varphi.
\end{displaymath}
For a negatively charged particle $\textbf{F}_Q$ is directed opposite to the electric field and thus opposite to $\textbf{F}_{\rm pol}$. The sum of two contributions is
\begin{equation}
{\mathcal R}_{\rm pol}+{\mathcal R}_{\rm Q} = \frac{Qe}{4\lambda_{\rm D}T_i}\left(1+\frac{J_0}{\Omega _{\rm ch}}\frac{e}{Q}\right).
\end{equation}

The factor $(1+\frac{n_{d0}}{n_{i0}}\frac{J_0}{\Omega_{\rm ch}})$ in the left-hand side of Eq.~(\ref{DR}) is associated with charge variations, it appears also in the case when both polarization and charge-gradient forces are neglected.\cite{FortovUFN}

Let us next compare the magnitudes of various terms in a special exemplary situation. We consider as an example a weakly collisional (low neutral gas pressure) gas discharge with electrons that are much hotter than ions. To describe particle charging in these conditions the orbital motion limited (OML) theory~\cite{AllenOML} is applicable. In this regime the relation  between charging frequency and ion/electron flux  (under the additional assumption $T_{e}\gg T_{i}$) is~\cite{FortovUFN}
\begin{equation}\label{Omega_ch}
\Omega_{\rm ch} \simeq J_0\frac{1+z}{z}\frac{e^2}{aT_e},
\end{equation}
where the reduced charge $z=|Q|e/aT_e$ has been introduced. The relative importance of polarization and charge gradient forces is
\begin{equation}\label{RatioR}
|{\mathcal R}_{\rm pol}/{\mathcal R}_{\rm Q}|\simeq (1+z)/4.
\end{equation}
Since typical values of $z$ are between $\simeq 2$ and $\simeq 4$ within the OML theory~\cite{FortovUFN,KhrapakPRE2001} this two component of the forces are of comparable magnitude. In the special case $z=3$, the two effects would completely cancel each other. In the weakly collisional regime the ion flux to the particle can be enhanced due to ion-neutral charge exchange collisions in the vicinity of the particle.\cite{ZobninJETP2000,LampePoP2003} As a results the charge tends to more positive values and the reduced charge $z$ can drop to values below unity.\cite{RatynskaiaPRL,KhrapakPRE2005,KhrapakEPL2012} This would indicate that CG contribution dominates. However, Eqs.~(\ref{charge2}), (\ref{Omega_ch}), and, hence, (\ref{RatioR}) should also be modified in this case. This regime would thus require careful additional consideration, which is beyond the scope of this brief communication.

Finally, we demonstrate that the factor in the left-hand side of Eq.~(\ref{DR}) associated with charge variations is normally close to unity in the considered case. Using Eq.~(\ref{Omega_ch}) it is easy to get
\begin{equation}
\frac{n_{d0}}{n_{i0}}\frac{J_0}{\Omega_{\rm ch}}=\frac{1}{1+z}\frac{|Z|n_{d0}}{n_{i0}}.
\end{equation}
The ratio $P_i=|Z|n_{d0}/n_{i0}$ (which can be termed the ion Havnes parameter) can approach unity only in rather extreme situation when all negative charge in the systems is residing on the particle component and the electron population is completely depleted.
Under more typical conditions, $P_i$ is well below unity and thus direct contribution from the charge variations to the real part of the dispersion relation is insignificant.

To conclude, we have investigated the effect of the charge-gradient force, associated with the charge variability in complex plasmas, on the propagation of low-frequency dust-acoustic waves. It has been demonstrated that the charge-gradient and polarization forces can be of comparable magnitude in collisionless plasmas with hot electrons, but act in the opposite directions. Charge-gradient effect can dominate for lower charges, while polarization effect becomes more important at higher charges. This should be properly taken into account when describing the dispersion of low-frequency dust acoustic waves in weakly coupled unmagnetized plasma.

 This work was supported by Presidium RAS program No.13 ``Condensed Matter and Plasma at High Energy Densities''. The work at  Aix-Marseille-University was supported by A*MIDEX project (Nr.~ANR-11-IDEX-0001-02) funded by the French Government ``Investissements d'Avenir'' program managed by the French National Research Agency (ANR).


\begin{thebibliography}{[1]}

\bibitem{TsytovichUFN} V.\,N. Tsytovich, \,Phys.-Usp. {\bf 40}, 53 (1997).
\bibitem{FortovUFN} V.\,E. Fortov, A.\,G. Khrapak, S.\,A. Khrapak, V.\,I. Molotkov, and O.\,F. Petrov, Phys.-Usp. {\bf 47}, 447 (2004).
\bibitem{ShuklaRMP} P.\,K. Shukla and B. Eliasson, Rev. Mod. Phys. {\bf 81}, 25 (2009).

\bibitem{Fortov05} V.\,E. Fortov, A. Ivlev, S. Khrapak, A. Khrapak, and G. Morfill, Phys. Rep. \textbf{421}, 1 (2005).
\bibitem{MorfillRMP} G.\,E. Morfill and A.\,V. Ivlev, Rev. Mod. Phys. {\bf 81}, 1353 (2009).

\bibitem{Rao90} N.\,N. Rao, P.\,K. Shukla, and M.\,Y. Yu, Planet. Space Sci. \textbf{38}, 543 (1990).
\bibitem{HamaguchiPRE1994} S. Hamaguchi and R.\,T. Farouki, Phys. Rev. E \textbf{49}, 4430,  (1994).
\bibitem{HamaguchiPoP1994} S. Hamaguchi and R.\,T. Farouki, Phys. Plasmas \textbf{1}, 2110 (1994).
\bibitem{KhrapakPRL2009} S.\,A. Khrapak, A.\,V. Ivlev, V.\,V. Yaroshenko, and G.\,E. Morfill, Phys. Rev. Lett. \textbf{102}, 245004 (2009).
\bibitem{BandiNJP2010} P. Bandyopadhyay, U. Konopka, S.\,A. Khrapak, G.\,E. Morfill, and A. Sen, New J. Phys. {\bf 12}, 073002 (2010).
\bibitem{BandiPoP2012} P. Bandyopadhyay, K. Jiang, R. Dey, and G.\,E. Morfill, Phys. Plasmas {\bf 19}, 123707 (2012).

\bibitem{MerlinoPPCF2012} R.\,L. Merlino, J.\,R. Heinrich, S.-H. Kim, and J.\,K. Meyer, Plasma Phys. Control. Fusion {\bf 54}, 124014 (2012).
\bibitem{PrajapatiPLA2015} R.\,P. Prajapati and S. Bhakta, Phys. Lett. A {\bf 379}, 2723 (2015).
\bibitem{SharmaEPL2016} P. Sharma and S. Jain, EPL {\bf 113}, 65001 (2016).
\bibitem{BentadetIEEE2017} K. Bentadet and M. Tribeche, IEEE Trans. Plasma Sci. {\bf 45}, 736 (2017).

\bibitem{KhrapakPRE03_2015} S.\,A. Khrapak and H.\,M. Thomas, Phys. Rev. E {\bf 91}, 033110 (2015).

\bibitem{Alexandrov84} A.\,F. Alexandrov, L.\,S. Bogdankevich, and A.\,A. Rukhadze,\emph{ Principles of Plasma Electrodynamics} (Springer, New York, 1984).

\bibitem{Fortov00} V.\,E. Fortov, A.\,G. Khrapak, S.\,A. Khrapak, V.\,I. Molotkov, A.\,P. Nefedov, O.\,F. Petrov,
and V.\,M. Torchinsky, Phys. Plasmas \textbf{7}, 1374 (2000).

\bibitem{AllenOML} J.\,E. Allen, Phys. Scr. {\bf 45}, 497 (1992).

\bibitem{KhrapakPRE2001} S.\,A. Khrapak, A.\,V. Ivlev, and G. Morfill, Phys. Rev. E {\bf 64}, 046403 (2001).

\bibitem{ZobninJETP2000} A.\,V. Zobnin, A.\,P. Nefedov, V.\,A. Sinel'shchikov, and V.\,E. Fortov, JETP {\bf 91}, 483 (2000).
\bibitem{LampePoP2003} M. Lampe, R. Goswami, Z. Sternovsky, S. Robertson, V. Gavrishchaka, G. Ganguli, and G. Joyce, Phys. Plasmas {\bf 10}, 1500 (2003).

\bibitem{RatynskaiaPRL}  S. Ratynskaia, S. Khrapak, A. Zobnin, M.\,H. Thoma, M. Kretschmer, A. Usachev, V. Yaroshenko, R.\,A. Quinn, G.\,E. Morfill, O. Petrov, and V. Fortov, Phys. Rev. Lett. {\bf 93}, 085001 (2004);

\bibitem{KhrapakPRE2005} S.\,A. Khrapak, S.\,V. Ratynskaia, A.\,V. Zobnin, A.\,D. Usachev, V.\,V. Yaroshenko, M.\,H. Thoma, M. Kretschmer, H. H\"{o}fner, G.\,E. Morfill, O.\,F. Petrov, and V.\,E. Fortov, Phys. Rev. E {\bf 72}, 016406 (2005).

\bibitem{KhrapakEPL2012} S.\,A. Khrapak, P. Tolias, S. Ratynskaia, M. Chaudhuri, A. Zobnin, A. Usachev, C. Rau, M.\,H. Thoma, O.\,F. Petrov, V.\,E. Fortov, and G.\,E. Morfill, EPL {\bf 97}, 35001 (2012).


\end{thebibliography}
\end{document}